\documentclass[pra,twocolumn,superscriptaddress]{revtex4-1}
\usepackage{graphicx,amsmath,mathdots}
\usepackage[colorlinks=true, citecolor=blue, urlcolor=blue,citecolor=blue]{hyperref}
\begin{document}
\title{Floquet engineering of localized propagation of light in waveguide array}
\author{Chao Ma}
\affiliation{School of Physical Science and Technology, Lanzhou University, Lanzhou 730000, China}
\author{Yuan-Sheng Wang}
\affiliation{School of Physical Science and Technology, Lanzhou University, Lanzhou 730000, China}
\author{Jun-Hong An}
\email{anjhong@lzu.edu.cn}
\affiliation{School of Physical Science and Technology, Lanzhou University, Lanzhou 730000, China}
\date{\today}

\begin{abstract}
The light propagating in a waveguide array or photonic lattice has become an ideal platform to control light and to mimic quantum behaviors in a classical system.  We here investigate the propagation of light in a coupled waveguide array with one of the waveguides periodically modulated in its geometric structure or refractive index. Within the framework of Floquet theory, it is interesting to find that the light shows the localized propagation in the modulated waveguide as long as bound quasistationary modes are formed in the band-gap area of the Floquet eigenvalue spectrum. This mechanism gives a useful instruction to confine light via engineering the periodic structure to form the bound modes. It also serves as a classical simulation of decoherence control via temporally periodic driving in open quantum systems.
\end{abstract}
\maketitle

\section{Introduction}
Dynamic localization, which generally involves electronic motion in a tight-binding crystalline lattice, means the suppression of quantum diffusion by an external potential. It includes Bloch oscillation induced by dc fields \cite{RevModPhys.34.645} and Anderson localization induced by disorder \cite{PhysRev.109.1492} in the static case, and dynamic localization induced by ac fields in the nonequilibrium case \cite{PhysRevB.34.3625,PhysRevLett.117.144104}. However, it is not easy to directly observe the dynamic localization of electrons in solid systems due to the severe decoherence effect of the electrons caused by the thermal oscillation of the lattices and correlation effect between electrons. It was found that the diverse quantum behaviors of electrons moving in the lattice can be classically simulated by the light tunneling in optical waveguide arrays \cite{Iyer:07,Chremmos:12,Segev2013,Block2014,PhysRevLett.114.243901,NEZHAD2016299}. Combined with the modern nanofabrication techniques of optical waveguides and photonic lattices in a highly clean level, this opens an avenue to directly visualize quantum behaviors in classical optical systems \cite{lederer2008discrete,LPOR:LPOR200810055,GARANOVICH20121}.

The formal similarity of the propagation equation of light in optical waveguide array to the Schr\"{o}dinger equation of electrons moving in a lattice builds the physical foundation to mimic quantum hopping of electrons by the optical system. Here the time evolution of the electron wave function is mapped to the propagation of the optical wave in coordinate space. The advantage of choosing lights instead of electrons is the versatility in engineering artificial optical potentials and its avoidance of the occurrence of decoherence and correlation effects of electrons. It makes the waveguide-based optical systems an ideal platform to simulate quantum effects \cite{PhysRevLett.96.023901,PhysRevLett.97.110402,PhysRevLett.110.076403,PhysRevLett.100.170506,Plotnik2014}. On the other hand, the controllability of light flowing is interesting on its own due to the numerous applications in optical engineering \cite{OptiEng2012}. Tremendous progress has been made in tailoring the light flowing in the spatially periodic structure. The optical Bloch oscillation has been proposed in the waveguide array by using linearly varying propagation constants to simulate the dc field \cite{Peschel:98,Efremidis:04,Chremmos:12,Xu2016}. It was observed in waveguide arrays \cite{PhysRevLett.83.4756,Block2014} even with periodic curvature along the propagation direction \cite{PhysRevLett.103.143903}. By introducing the disorder to the propagation constant, Anderson localization has been observed in disordered photonic lattices \cite{PhysRevLett.100.013906,Schwartz2007,Segev2013}. The dynamic localization induced by an ac field can also be optically simulated. It was found that the light propagating in a parallelly separated waveguide array with all the waveguides periodically curved shows the localization \cite{Iyer:07,PhysRevLett.102.153901,PhysRevLett.104.223903,PhysRevLett.109.103901,SzameitNP2009,NEZHAD2016299}. The dynamic localization due to such periodic curvature was attributed to a resonance effect that can only occur for certain discrete values of the ratio between the amplitude and the frequency of the periodic curvature to the waveguides \cite{PhysRevLett.102.153901,SzameitNP2009}. It raises a high requirement to the fabrication technique to curve the waveguides. Note that the resonance condition is obtained by neglecting all higher-order terms of the Bessel function expansions to the exponential of the periodic curvature function in the high-frequency-modulation condition.

In this work, we study the propagation of light in a waveguide array with only one waveguide periodically modulated either to its geometric curvature or to its refractive index under the framework of Floquet theory. A similar structure \cite{PhysRevLett.101.143602} was used to simulate the control of quantum-mechanical decay under the Markovian approximation in the weak-coupling condition \cite{PhysRevLett.87.270405}. Due to the spatial dependence introduced by the periodic modulation, the light propagating in such structure has no well-defined propagation constants and stationary modes. Thanks to Floquet theory, it can be well characterized by the quasistationary modes determined by the Floquet eigenequation with the corresponding eigenvalues acting as the propagation constant. A mechanism of the localized propagation of light in the modulated waveguide is found from the point of view of the Floquet eigenvalue spectrum. Going beyond the weak-coupling condition, we find that the light in the principal waveguide during the propagation tends to vanish if only a continuous band is present in the eigenvalue spectrum; whereas, it tends to be preserved if isolated Floquet bound modes (FBMs) are formed in the gap area of the eigenvalue spectrum. Different from the Markovian approximate result in Ref. \cite{PhysRevLett.101.143602}, where the decay of the light strength in the waveguide with the periodic modulation can only be slowed down or sped up, our result indicates that the light spreading can be totally suppressed. Compared with the resonance mechanism of the dynamic localization in Ref. \cite{PhysRevLett.102.153901,SzameitNP2009}, it is several wide bands of the modulation parameters where the light shows the localized propagation. This implies that our mechanism is robust to the inevitable fluctuation of the modulation parameters, which hopefully could simplify the fabrication technique in designing the modulation structure.

This paper is organized as follows. In Sec. \ref{owd}, we give the model of the light propagating in a coupled waveguide array based on the coupled mode theory. Section \ref{os} focuses on the localized propagation of light when one of the waveguides is periodically modulated. The mechanism of the localized propagation is attributed to the formation of the FBMs. Finally, a summary is given in Sec. \ref{sum}.

\begin{figure}[tbp]
\centering\includegraphics[width = .95 \columnwidth]{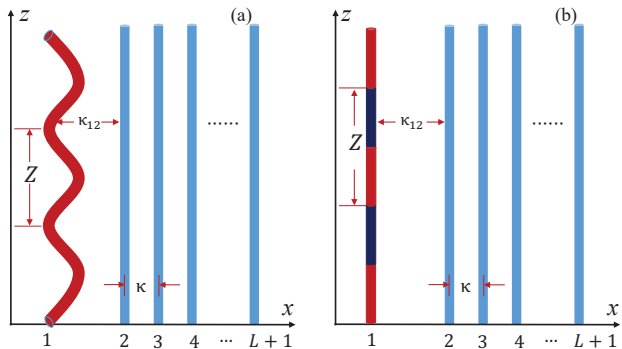}
\caption{Schematic diagram of light prorogating in a waveguide periodically modulated in its geometric structure (a) and composite materials (b) coupled to a waveguide array. $\kappa_{12}$ is the coupling strength of light between the principal waveguide and the first one of the array, $\kappa$ is the nearest-neighbor coupling strength of the light in the array, and $Z$ is the modulation period.}\label{scheme}
\end{figure}
\section{Light propagating in a coupled waveguide array}\label{owd}
We consider the physical setting in which a single-mode optical field polarized in the $x$ direction propagates along the $z$ direction in $L+1$ optical waveguides arrayed in a one-dimensional configuration. The light is initially injected into the first waveguide acting as the principal waveguide by a focused laser beam. The overlap between the optical evanescent modes in the waveguides allows the propagating light to tunnel into its neighboring waveguides and consequently the light spreads throughout the array. In the paraxial approximation, the propagation equation of the field in the waveguide array is governed by \cite{Jones:65,GARANOVICH20121,PhysRevLett.83.4756,Rechtsman2013}
\begin{equation}\label{ou}
    i\lambdabar{d\mathbf{A}(z)\over dz}=\mathbf{M}\mathbf{A}(z),
\end{equation}
where $\lambdabar=\lambda/2\pi$ with $\lambda$ the wavelength of the light, $\mathbf{A}(z)=\left(
                       \begin{array}{cccccc}
                         A_1(z) & A_2(z) &  A_3(z)& \cdots &  A_{L+1}(z) \\
                       \end{array}
                     \right)^T $ under the condition $\mathbf{A}(0)=\left(
                       \begin{array}{cccccc}
                         1 & 0 & \cdots &  0 \\
                       \end{array}
                     \right)^T $ denote the light amplitudes in the waveguides, and
\begin{equation}
\mathbf{M}=\left(\begin{array}{cccccc}
\lambdabar\beta_1 & \kappa_{12} & 0 & \cdots &  0 \\
\kappa_{21} & \lambdabar\beta & \kappa & \cdots &  0 \\
0 & \kappa & \lambdabar\beta & \cdots & 0 \\
\vdots & \vdots & \vdots & \ddots & \vdots\\
0 & 0 & 0 & \dots &  \lambdabar\beta \\
\end{array}\right).\label{MMM}
\end{equation}
Here we have assumed the propagation constants and the nearest neighboring coupling strength of the light in the waveguide array are identical, i.e. $\beta_2=\cdots=\beta_{L+1}\equiv \beta$ and $\kappa_{32}=\cdots=\kappa_{L+1,L}\equiv\kappa$, under the condition that the $L$ waveguides in the array are identical and equally separated. The nearest neighboring coupling in the waveguide array defines a dispersion relation $\beta_k=\beta+2\kappa\cos (k)/\lambdabar$ with $k$ the Bloch wave vectors, which results in a single continuous band with width $4\kappa/\lambdabar$ to the propagation constants of the light in the waveguide array. It resembles the tight-binding model in describing quantum transport of electrons in crystalline lattices \cite{PhysRevLett.75.3914,Block2014}. Such resemblance makes the classical optical field propagating in a waveguide array an ideal system to simulate the diverse quantum effects of electrons in crystalline lattices.

\section{Localized propagation of light by periodic modulation}\label{os}
Further consider that the principal waveguide is periodically modulated in its structure such that its propagation constant $\beta_1$ or coupling strength $\kappa_{12}$ is $z$-dependent with period $Z$ (see Fig. \ref{scheme}). Then the coefficient matrix in Eq. (\ref{MMM}) changes into $z$-dependent, too. Besides directly solving via numerical calculation, the solution of Eq. (\ref{ou}) with $\mathbf{M}(z)=\mathbf{M}(z+Z)$ can be determined by the Floquet theorem \cite{PhysRev.138.B979,PhysRevA.7.2203,PhysRevA.95.023615}
\begin{equation}
\mathbf{A}(z)=\sum_\alpha c_\alpha e^{-i\epsilon_\alpha z}u_\alpha (z),\label{Azz}
\end{equation}
where $c_\alpha =u_\alpha (0)^T\cdot\mathbf{A}(0)$ only depends on the initial condition, $\epsilon_\alpha $ and $u_\alpha (z)=u_\alpha (z+Z)$ satisfy
\begin{equation}
[\mathbf{M}(z)-i\lambdabar\partial_z]u_\alpha (z)=\lambdabar\epsilon_\alpha u_\alpha (z).\label{flqt}
\end{equation}
The independence of $c_\alpha $ on $z$ means that $u_\alpha (z)$ and $\epsilon_\alpha $ play the same role as the optical stationary modes and propagation constants in the static system without the periodic modulation, respectively. Thus they are called quasistationary modes and quasipropagation constants. By a Fourier transform $u_\alpha (z)=\sum_m e^{im\omega z}\tilde{u}_\alpha(m)$ with $\omega=2\pi/Z$, Eq. (\ref{flqt}) is recast into
\begin{equation}
\sum_m [\tilde{\mathbf{M}}_{n-m}+m\lambdabar\omega\delta_{m,n}]\tilde{u}_\alpha(m)=\lambdabar\epsilon_\alpha \tilde{u}_\alpha(n),\label{flqel}
\end{equation}
where $\tilde{\mathbf{M}}_{n-m}=Z^{-1}\int_0^Z\mathbf{M}(z)e^{-i(n-m)\omega z}dz$. Then expanding each $\tilde{\mathbf{M}}_{n-m}$ in the complete basis of the $L+1$ waveguides, we get an infinite-rank matrix equation. The quasistationary modes are obtained by truncating the basis to the rank such that the obtained magnitudes converge. Note that $e^{-in\omega z}u_\alpha(z)$ is also the eigenstate of the Floquet eigenequation (\ref{flqt}) with the eigenvalue $\lambdabar(\epsilon_\alpha+n\omega)$. Thus the eigenvalues are periodic with period $\omega$ and one generally chooses them within $[-\omega/2,\omega/2]$ called the first Brillouin zone.

First, we study that the principal waveguide is harmonically modulated in its geometric structure [see Fig. \ref{scheme}(a)]. It is based on a curved waveguide in which the curvature varies periodically with propagation distance $z$. This structure and its extension to a two-dimensional configuration have been used to simulate quantum Zeno effects \cite{PhysRevLett.97.110402}, decoherence control \cite{PhysRevLett.101.143602}, and photonic Floquet topological insulator \cite{Rechtsman2013,Maczewsky2017}. Such required periodic bending profile can be realized via the laser direct-writing method in fused-silica glass \cite{SzameitNP2009}. The distance between the adjacent sites of waveguides in the $x$ direction characterizes the coupling strength. Thus, the modulation causes
\begin{equation}
\kappa_{12}(z)=a\cos(\omega z)+b,\label{kap12}
\end{equation}
where $a$ and $\omega$ are the amplitude and frequency of the periodic modulation, respectively, and $b$ is a constant shift.

\begin{figure}[tbp]
\centering\includegraphics[width = 1.0 \columnwidth,clip]{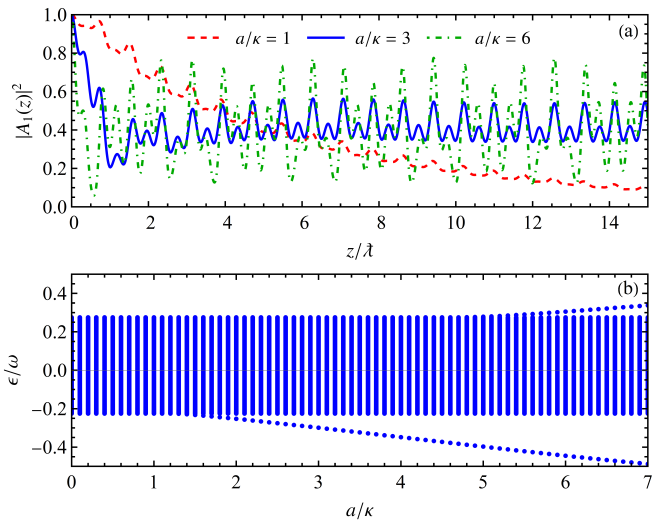}
\caption{(a) Propagation of the light strength $|A_1(z)|^2$ in the principal waveguide in different modulation amplitude $a$ obtained by numerically solving Eq. (\ref{ou}) with the periodic $\kappa_{12}$ (\ref{kap12}). (b) Spectrum of Floquet eigenvalues in different modulation amplitude $a$ obtained by numerically solving Eq. (\ref{flqel}). The parameters are $b=0.6\kappa$, $\beta_{1}=6.5\lambdabar^{-1}$, $\beta=0.2\lambdabar^{-1}$,~$\omega=8.0\lambdabar^{-1}$, and $L=200$.}\label{prooverall}
 \end{figure}
Solving the propagation equation (\ref{ou}) with the $z$-dependent $\kappa_{12}$ in Eq. (\ref{kap12}) numerically, we can obtain the evolution of the amplitude of light in the whole waveguides with the propagation distance $z$. Figure \ref{prooverall}(a) shows the propagation of the light strength $|A_{1}(z)|^2=|\mathbf{A}(0)^T\cdot\mathbf{A}(z)|^2$ of the principal waveguide in different modulation amplitude $a$. It can be found that $|A_{1}(z)|^2$ decays monotonically to zero and the light in the principal waveguide spreads completely to the waveguide array for small modulation amplitude $a$. With the increase of $a$, the light in the principal waveguide tends to lossless oscillation and its decay is efficiently suppressed. In Appendix \ref{comspct}, we make a detailed comparison of our exact result with the one obtained using the Markovian approximation under the weak-coupling limit \cite{PhysRevLett.101.143602}, which shows the substantial difference. Our exact result reveals the localized propagation of light induced by the periodic modulation, while the approximate one indicates that the decay of the light is only slowed down or sped up. It demonstrates that the periodic modulation not only can change quantitatively the propagation behavior of light, but also can change qualitatively its steady-state behavior. This supplies a constructive idea to control the tunneling of light in the waveguides.

\begin{figure}[tbp]
\centering\includegraphics[width = \columnwidth,clip]{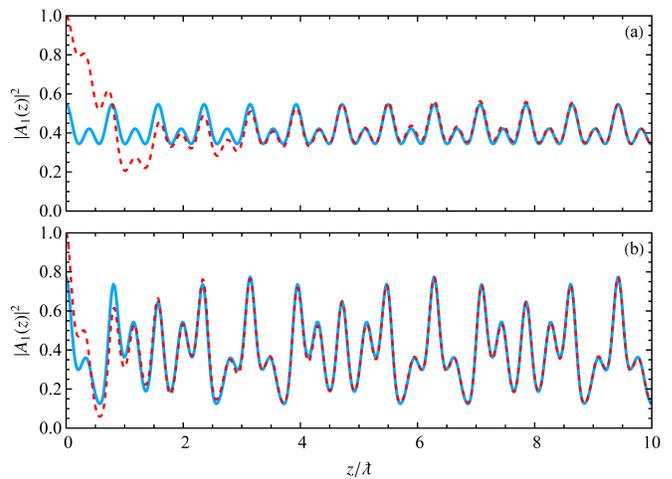}
\caption{$|A_1(z)|^2$ calculated by numerically solving Eq. (\ref{ou}) (red dashed lines) and by evaluating the analytical solution (\ref{asympt}) (blue solid lines) in one FBM case when $a/\kappa=3$ (a) and in two FBM case when $a/\kappa=6.5$ (b). Other parameters are the same as Fig. \ref{prooverall}.}\label{comparison}
\end{figure}
To understand the physical mechanism of the localized propagation induced by the periodic modulation, we resort to the Floquet theorem. Figure \ref{prooverall}(b) plots the spectrum of the Floquet eigenvalues with the change of the modulation amplitude $a$ via numerically solving Eq. (\ref{flqel}). It indicates that the region where the light shows the localized propagation matches well with the one where isolated bound quasistationary modes are formed in the gap area of the spectrum. We call such isolated bound quasistationary modes FBMs. Different from the widely studied bound mode in the continuum \cite{PhysRevLett.107.183901,Longhi2013,PhysRevA.93.062122,BIC2016}, our FBM is an optical bound quasistationary mode residing within the gap area in the periodically modulated waveguide system. It is interesting to see that the light strength in the principal waveguide tends to lossless oscillation as long as the FBMs are formed. To illustrate this behavior, we, according to the Floquet theorem (\ref{Azz}), can construct the solution of Eq. (\ref{ou}) with the $z$-dependent $\mathbf{M}(z)$ as
\begin{equation}
\mathbf{A}(z)=\sum_{l=1}^M d_{0,l} e^{-i\epsilon_{0,l} z}u_{0,l} (z)+\sum_{\alpha\in\text{Band}} c_\alpha e^{-i\epsilon_\alpha z}u_\alpha (z), \label{expan}
\end{equation}
where $d_{0,l}=u_{0,l}(0)^T\cdot \mathbf{A}(0)$ and $M$ is the number of the formed FBMs. The two terms in Eq. (\ref{expan}) correspond to the contributions from the $M$ potentially formed FBMs and the modes in the continuous band of the spectrum, respectively. To our nearest-neighboring tight-binding model, only a single continuous band can be formed. Therefore, at most two FBMs can be formed. Oscillating with $z$ in continuously changing frequencies $\epsilon_\alpha$, the second term in Eq. (\ref{expan}) behaves as a decay and tends to zero due to out-of-phase interference of the terms in the continuous bands. Then the asymptotic solution of the principal waveguide reads
\begin{equation} \label{asympt}
|A_1(\infty)|^2=\left\{ \begin{aligned}
         &0,\hspace{4.2cm}M=0\\
         &|d_0\mathbf{A}(0)^T\cdot u_0(z)|^2,\hspace{1.6cm}M=1 \\
         &\sum_{l=1}^2 |d_{0,l} \mathbf{A}(0)^T\cdot u_{0,l} (z)|^2+D,M=2
                          \end{aligned} \right.
\end{equation}
with $D=2\cos[(\epsilon_{0,1} -\epsilon_{0,2} )z]\otimes_ld_{0l}\mathbf{A}(0)^T\cdot u_{0,l}(z)$ denoting the interference between the two FBMs.

Figure \ref{comparison} depicts the comparison of $|A_1(z)|^2$ obtained by numerically solving Eq. (\ref{ou}) and by evaluating the analytical solution (\ref{asympt}). It shows that the numerical result, after a small jolt in the small-$z$ regime, coincides exactly with the analytical result in Eq. (\ref{asympt}) in the large-$z$ limit. When only one FBM is formed, $|A_1(\infty)|^2$ shows the perfect oscillation with the same frequency $\omega$ as the FBM $u_0(z)$ [see Fig. \ref{comparison}(a)]. It demonstrates that the light coherently synchronizes with the periodic modulation in this single FBM case. When two FBMs are formed, $|A_1(\infty)|^2$ shows the lossless oscillation with multiple frequencies jointly determined by $\omega$ and $\epsilon_{0,1} -\epsilon_{0,2}$ due to the interference between the two FBMs [see Fig. \ref{comparison}(b)]. The comparison unambiguously validates the FBM mechanism in governing the optical localized propagation induced by the periodic modulation.

\begin{figure}[tbp]
\centering
\includegraphics[width = \columnwidth,clip]{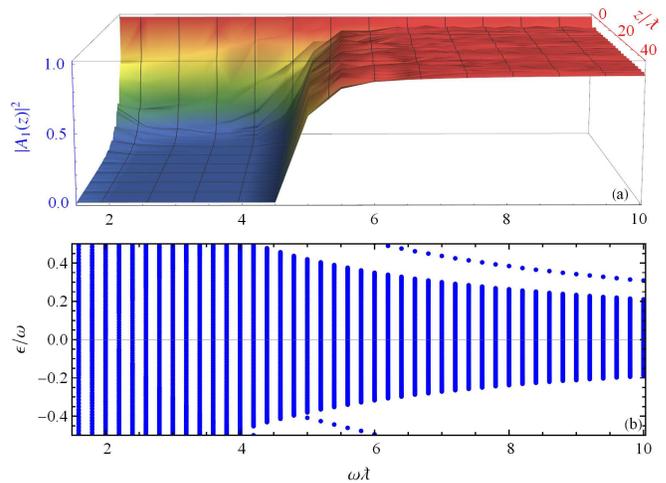}
\caption{(a) Propagation of the light strength $|A_1(z)|^2$ in the principal waveguide in different modulation frequency $\omega$ obtained by numerically solving Eq. (\ref{ou}) with the periodic $\kappa_{12}$ (\ref{kap12}). (b) Spectrum of Floquet eigenvalues in different modulation frequency $\omega$ obtained by numerically solving Eq. (\ref{flqel}). The parameters are $a=1.0\kappa$, $b=0.5\kappa$, $\beta_{1}=3.0\lambdabar^{-1}$, $\beta=0.1\lambdabar^{-1}$, and $L=200$.}
\label{freqc}
 \end{figure}
The above result reveals that we can manipulate the spectrum of the Floquet eigenvalues forming the FBM to suppress the spread of light in the waveguide array. A prerequisite for forming the FBM is the existence of a finite gap in the spectrum. We plot in Fig. \ref{freqc} $|A_1(z)|^2$ and the Floquet eigenvalue spectrum with the change of the modulation frequency $\omega$. Once again it confirms the firm correspondence between the formation of the FBM and the localized propagation of light. Different from Fig. \ref{prooverall}, the width of the band gap is not constant with the change of the modulation frequency in this case. This can be understood in the following way. Periodic in $\omega$, the Floquet spectrum has a full width $\omega$. The width of the continuous band of the propagation constant is $4\kappa/\lambdabar$. Therefore, a band gap with width $\omega -4\kappa/\lambdabar$ can be present in the spectrum only in the high-frequency (i.e., $\omega > 4\kappa/\lambdabar$) modulation case. This is indeed verified by Fig. \ref{freqc}(b) where the band gap vanishes whenever $\omega < 4\kappa/\lambdabar=4.0/\lambdabar$. It leads to the continuous band of the waveguide filling up the whole Floquet spectrum. Thus no room would be left for forming the FBM here. Reflecting on $|A_1(z)|^2$ in Fig. \ref{freqc}(a), $|A_1(z)|^2$ approaches zero eventually in the low-frequency case. Therefore, we conclude that the FBM can be present only when the modulation frequency satisfies $\omega > 4\kappa/\lambdabar$, which supplies a necessary condition to realize the localized propagation of light. It is a very useful criterion on designing a modulation scheme for controlling the optical tunneling in the waveguide array.

\begin{figure}[tbp]
\centering
\includegraphics[width = 1.0 \columnwidth,clip]{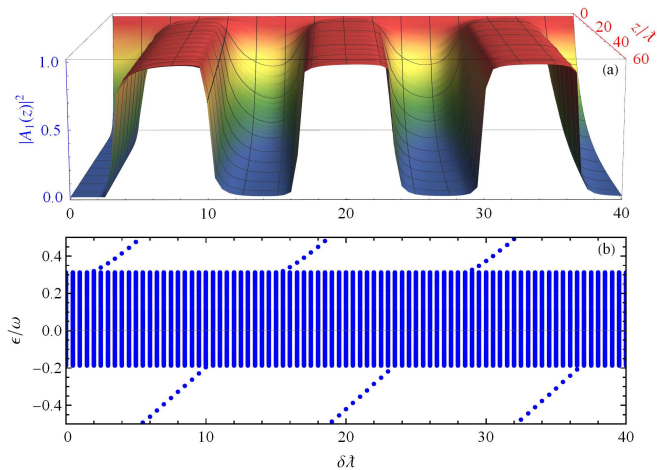}
\caption{Propagation of the light strength $|A_1(z)|^2$ in the principal waveguide (a) and spectrum of Floquet eigenvalues (b) in different modulation amplitude $\delta$ to the propagation constant $\beta_1$. The parameters are $Z=0.25\pi\lambdabar$,~$Z'=0.1\pi\lambdabar$,~$\kappa_{12}=0.5\kappa$, $\beta=\beta_0=0.5\lambdabar^{-1}$, and $L=200$.}
\label{Figure:Pluse1}
\end{figure}

Second, we consider that the linear refractive index of the principal waveguide along the propagation direction is periodically modulated in step function [see Fig. \ref{scheme}(b)]. This can be readily realized by the periodic fabrication with different composite materials in the waveguide. It causes
\begin{equation} \label{stepd}
\beta_1(z)=\left\{ \begin{aligned}
         \beta_0,~&z\in[nZ,nZ+Z'] \\
         \beta_0+\delta,~&z\in[nZ+Z',(n+1)Z]
                          \end{aligned} \right..
                          \end{equation}
Different from Refs. \cite{PhysRevLett.102.153901,SzameitNP2009}, which focused on deriving the resonance mechanism of the dynamic localization by neglecting the higher-order terms of Bessel function expansion to the exponential of the periodic modulation, we here give an exact study to the propagation dynamics of the light in such modulated waveguide structure. Figure \ref{Figure:Pluse1} shows the propagation of light strength $|A_1(z)|^2$ and the spectrum of Floquet eigenvalues with the change of the modulation amplitude $\delta$. It reveals again that the tunneling of light is suppressed whenever the FBM is formed in the spectrum of Floquet eigenvalues. Another interesting conclusion that can be drawn is that it is a wide band instead of isolated points \cite{PhysRevLett.102.153901,SzameitNP2009} of the modulation amplitude $\delta$ that can cause the localized propagation. This makes our FBM mechanism to realize light localization robust to the imperfect fluctuation of the modulation parameters. To illustrate the difference between our FBM mechanism and the conventional dynamic localization \cite{PhysRevLett.102.153901,SzameitNP2009}, we make a detailed comparison to the results obtained by the two methods in Appendix \ref{comcdl}.

\section{Conclusion}\label{sum}
In summary, within the framework of Floquet theory, we have studied the propagation of light in the coupled waveguide array under the periodic modulation either to the geometric structure or to the refractive index on one of the waveguides. It is revealed that the mechanism of the localized propagation of light induced by the periodic modulation is the formation of isolated bound quasistationary modes in the bandgap area of the Floquet eigenvalue spectrum. Our finding supplies a way to control the light propagation in the waveguide array via well-tailored material fabrication to engineer the formation of the bound quasistationary modes. The mechanism can also be seen as a classical analog of the Floquet-bound-state-induced decoherence suppression in a quantum spin system \cite{PhysRevA.91.052122}. Given the wide application of a modulated waveguide array or photonic crystal in optical engineering and in classical simulation of quantum effects, our result may supply instructive insight to control and bend light in a structured medium. Note that other modulation structures to the waveguide array may be considered, but the main results remain qualitatively similar. It is also noted that our mechanism of the light localization does not incorporate with the optical Kerr nonlinearity \cite{PhysRevA.78.031801}, which deserves further exploration.

\section*{Acknowledgments}
This work is supported by the National Natural Science Foundation (Grant No. 11474139) and by the Fundamental Research Funds for the Central Universities of China.

\appendix
\section{Comparison with the spectral filtering method}\label{comspct}
The spectral filtering method was originally put forward in controlling the quantum-mechanical decay via time-dependent modulation \cite{PhysRevLett.87.270405}. It has been simulated in a classical optical system \cite{PhysRevLett.101.143602}. However, this method was established on the Markovian approximation under the weak-coupling condition. To demonstrate the difference between this method and ours, we in the following study our system using the spectral filtering method.
\begin{figure}[tbp]
\centering\includegraphics[width = 1 \columnwidth]{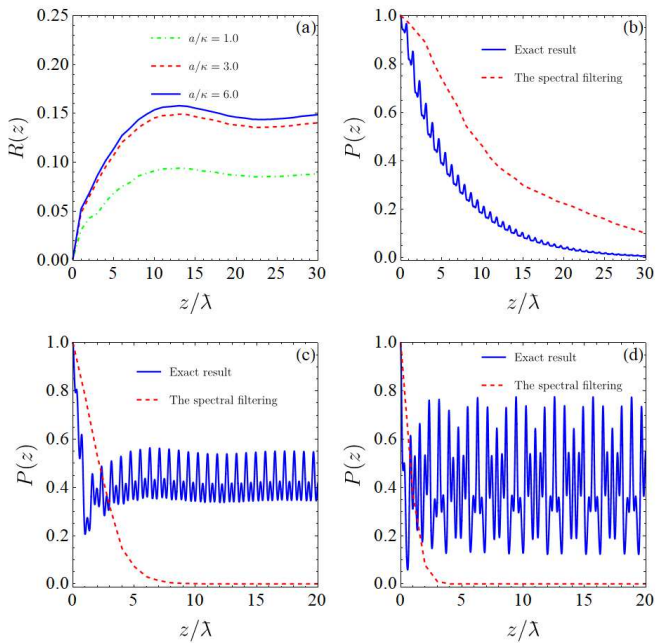}
\caption{(a) Damping rates of the light strength in the principal waveguide calculated from the spectral filtering method. Comparison of $P(z)$ calculated from the spectral filtering method [i.e., $P(z)=|\alpha(z)|^2$] (the red dotted lines) and from our exact method (the solid blue lines) when $a/\kappa=1.0$ in (b), 3.0 in (c), and 6.0 in (d). Other parameters are the same as those in Fig. \ref{prooverall}(a). }\label{cos}
\end{figure}
\begin{figure}[tbp]
\centering\includegraphics[width = 1 \columnwidth]{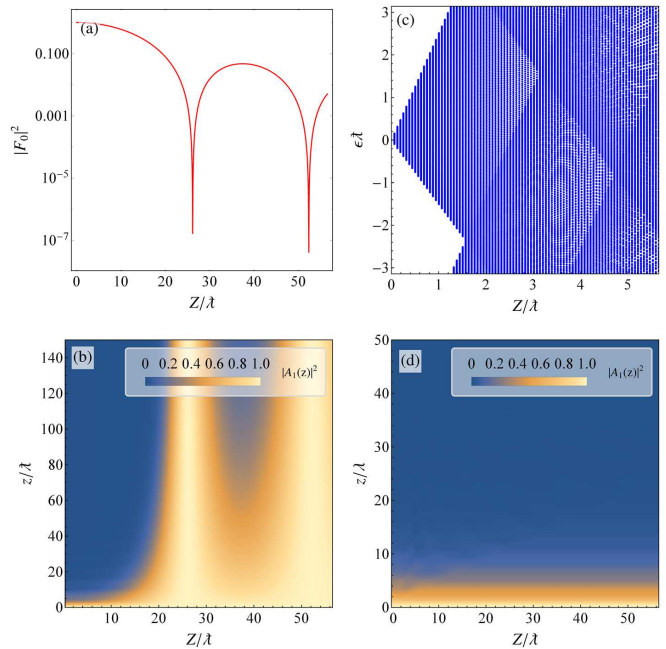}
\caption{Renormalized factor $|F_0|^2$ (a) and propagation of $|A_1(z)|^2$ (b) in different modulation period $Z$ calculated from the method of the conventional dynamic localization. Spectrum of Floquet eigenvalues (c) and exact propagation of $|A_1(z)|^2$ (d) in different modulation period $Z$ calculated from our exact treatment. The parameters are $Z'=0.4Z$, $\beta_0=\beta=0.5\lambdabar^{-1}$, $\delta=1.0\lambdabar^{-1}$, and $\kappa_{12}=0.5\kappa$. The result of $Z/\lambdabar>6$ in (c) is not shown because it is the same as that of $1.5\lesssim Z/\lambdabar<6$.  }\label{compps}
\end{figure}

From Eq. (\ref{ou}) with $z$-dependent $\kappa_{12}(z)$, we can derive the propagation equation of the light amplitude in the principal waveguide as
\begin{eqnarray}
&&{dA_{1}(z)\over dz}+i\beta_{1}A_{1}(z)\nonumber \\&&+\int^{z}_{0}dz'\kappa_{12}(z)\kappa_{12}(z')f(z-z')A_{1}(z')dz'=0,\label{AA1}
\end{eqnarray}
where $f(z-z')=\int d\beta G(\beta)e^{-i\beta(z-z')}$, $G(\beta)=\sum_{k}|g_{k}|^{2}\delta(\beta-\beta_{k})$ with $g_{k}=e^{-ik}/(\lambdabar\sqrt{N})$.
Under the weak coupling condition, we can make the Markovian approximation to Eq. (\ref{AA1}) and obtain
\begin{equation}\dot{\alpha}(z)\approx-\alpha(z)\int^{z}_{0}dz'\kappa_{12}(z)\kappa_{12}(z')f(z-z')e^{i\beta_{1}(z-z')}dz',\end{equation}
with $\alpha(z)=e^{i\beta_{1}z}A_{1}(z)$. Its solution can be readily obtained as \cite{PhysRevLett.87.270405,PhysRevLett.101.143602}
\begin{equation}
|\alpha(z)|=\exp[-\frac{1}{2}R(z)Q(z)],
\end{equation}
where $R(z)=2\pi\int^{\infty}_{-\infty}d\beta G(\beta+\beta_{1})|D_{z}(\beta)|^{2}/Q(z)$ with $D_{z}(\beta)=\frac{1}{\sqrt{2\pi}}\int^{z}_{0}e^{i\beta z'}\kappa_{12}(z')dz'$ and $Q(z)=\int^{z}_{0}|\kappa_{12}(z')|^{2}$ acts as a damping rate. Thus, under the Markovian approximation, the damping of the light amplitude in the principal waveguide is determined by the overlap integral between the noise spectrum and the spectrum of the periodic modulation. This is the physical meaning of the spectral filtering method.

We plot in Fig. \ref{cos} the comparison of our exact result with the one obtained from the spectral filtering method using the parameters in Fig. \ref{prooverall}. Figure \ref{cos}(a) indicates that in all three cases the noise spectrum and the spectrum of the periodic modulation have a notable overlap and thus the damping rates have nonzero values. Consequently, the light in the principal waveguide decays to zero without exception [see the red dashed lines in Figs. \ref{cos}(b)-\ref{cos}(d)]. However, our exact result shows that this is the truth only when the FBM is absent in the small modulation amplitude case. With the formation of the FBMs in the large modulation amplitude region, a qualitative difference of our exact result from the one obtained by the spectral filtering method can be observed [see Figs. \ref{cos}(c) and \ref{cos}(d)]. Therefore, it is the formed FBMs which is the physical reason for localized propagation of light in the periodically modulated waveguide array. The comparison reveals that, although the Markovian approximation used in the spectral filtering method could capture some propagation behavior of light, it might miss the correct physics in certain parameter regimes.

\section{Comparison with the conventional dynamic localization}\label{comcdl}
The difference of our result from the conventional dynamic localization is reflected not only from the system, where only one waveguide is modulated in ours, while all the waveguides are modulated in the conventional one, but also from the physical mechanism. The conventional dynamic localization of light is attributed to the resonant localization due to a periodic modulation of the system \cite{PhysRevLett.102.153901,SzameitNP2009}. This is reminiscent of the coherent destruction of tunneling \cite{Grifoni1998,Luo2014} and dynamical decoupling \cite{Zhou2009} in a quantum system induced by periodic driving. Such resonance and decoupling are obtained by neglecting the higher-order terms of the Bessel function expansions to the exponential of the periodic curvature function in the high-frequency-modulation condition. To demonstrate the difference between our mechanism and the conventional one, we here compare our result with that obtained from the conventional method by taking the modulation scheme (\ref{stepd}) as an example.

Expanding $u_\alpha(z)$ in Eq. (\ref{flqt}) in the basis of a rotating frame as $u_\alpha(z)=\sum_m \mathbf{U}_ze^{im\omega z}\tilde{u}_\alpha(m)$ with $\mathbf{U}_z=\exp[-i\mathbf{M}_0\int_0^z(\beta_1(z')-\bar{\beta}_1)dz']$, $\mathbf{M}_0=\left(
                \begin{array}{cccc}
                  1 & 0 & \cdots & 0 \\
                  0 & 0 & \cdots & 0 \\
                  \vdots & \vdots & \ddots & \vdots \\
                  0 & 0 & 0 & 0 \\
                \end{array}
              \right)
$, and $\bar{\beta}_1=Z^{-1}\int_0^Z\beta_1(z)dz$, we can obtain a similar form as Eq. (\ref{flqel}) with
\begin{widetext}\begin{eqnarray}
\tilde{\mathbf{M}}_{n-m}=\delta_{n,m}\left(\begin{array}{cccccc}
\lambdabar(\bar{\beta}_1+\beta_0) & 0 & 0 & \cdots &  0 \\
0 & \lambdabar\beta & \kappa & \cdots &  0 \\
0 & \kappa & \lambdabar\beta & \cdots & 0 \\
\vdots & \vdots & \vdots & \ddots & \vdots\\
0 & 0 & 0 & \dots &  \lambdabar\beta \\
\end{array}\right)+\left(\begin{array}{cccccc}
0 & \kappa_{12}F_{n-m} & 0 & \cdots &  0 \\
\kappa_{12}F_{n-m} & 0 & 0 & \cdots &  0 \\
0 & 0 & 0 & \cdots & 0 \\
\vdots & \vdots & \vdots & \ddots & \vdots\\
0 & 0 & 0 & \dots & 0 \\
\end{array}\right)
\end{eqnarray}\end{widetext}
with $F_{n-m}=Z^{-1}\int_0^Z\exp[-i\int_0^z(\beta_1(z')-\bar{\beta}_1)dz'-i(n-m)\omega z]dz$. In the high-frequency modulation condition, we neglect all the fast oscillating terms with $ n \neq m$ in $F_{n-m}$ and only keep $F_0$ in the same sprit of the rotating-wave approximation in quantum optics. Then the system is reduced to a static one with the coupling strength between the principal waveguide and its nearest-neighboring waveguide renormalized by a factor $F_0$ and the propagation constant shifted by a constant $\bar{\beta}_1$. Therefore, the light localization can be readily realized by manipulating the periodic modulation such that $F_0=0$.

From Fig. \ref{compps}(a), we can see that we really can realize $F_0=0$ at certain resonance points of the modulation parameter $Z$. This makes the light in the principal waveguide effectively decoupled to the waveguide array. As expected, the light shows the dynamic localization at the points with $F_0=0$ [see Fig. \ref{compps}(b)]. However, our exact calculation indicates that no localized propagation of light is present at these resonance points [see Fig. \ref{compps}(d)] because no FBM is formed in the whole parameter regime  [see Fig. \ref{compps}(c)]. It demonstrates that the approximation used in obtaining resonance decoupling in the conventional dynamic localization cannot govern all the physics of our studied model. It also differentiates our mechanism on the light localization from the conventional dynamic localization.

\bibliography{Ref}
\bibliographystyle{apsrev4-1}
\end{document}